\documentclass[preprint2]{aastex}

\usepackage{amsmath}
\usepackage{graphicx}
\usepackage{epsfig,psfrag,epic,eepic}
\usepackage{textcomp}
\usepackage{times}
\usepackage{longtable}

\slugcomment{Accepted to the Astrophysical Journal}

\shorttitle{Imaging Survey of Debris Disk Systems}
\shortauthors{Janson et al.}

\begin{document}

\title{The SEEDS Direct Imaging Survey for Planets and Scattered Dust Emission in Debris Disk Systems}

\author{Markus Janson\altaffilmark{1,28}, 
Timothy~D. Brandt\altaffilmark{1}, 
Amaya Moro-Mart\'in\altaffilmark{2}, 
Tomonori Usuda\altaffilmark{3}, 
Christian Thalmann\altaffilmark{4}, 
Joseph~C. Carson\altaffilmark{5,14}, 
Miwa Goto\altaffilmark{6}, 
Thayne Currie\altaffilmark{7}, 
M.~W. McElwain\altaffilmark{8}, 
Yoichi Itoh\altaffilmark{9}, 
Misato Fukagawa\altaffilmark{10}, 
Justin Crepp\altaffilmark{11}, 
Masayuki Kuzuhara\altaffilmark{12}, 
Jun Hashimoto\altaffilmark{12}, 
Tomoyuki Kudo\altaffilmark{3}, 
Nobuhiko Kusakabe\altaffilmark{12}, 
Lyu Abe\altaffilmark{13}, 
Wolfgang Brandner\altaffilmark{14}, 
Sebastian Egner\altaffilmark{3}, 
Markus Feldt\altaffilmark{14}, 
Carol~A. Grady\altaffilmark{15}, 
Olivier Guyon\altaffilmark{3}, 
Yutaka Hayano\altaffilmark{3}, 
Masahiro Hayashi\altaffilmark{3}, 
Saeko Hayashi\altaffilmark{3}, 
Thomas Henning\altaffilmark{14}, 
Klaus~W. Hodapp\altaffilmark{16}, 
Miki Ishii\altaffilmark{3}, 
Masanori Iye\altaffilmark{12}, 
Ryo Kandori\altaffilmark{12}, 
Gillian~R. Knapp\altaffilmark{1}, 
Jungmi Kwon\altaffilmark{17}, 
Taro Matsuo\altaffilmark{18}, 
Shoken Miyama\altaffilmark{19}, 
Jun-Ichi Morino\altaffilmark{12}, 
Tetsuro Nishimura\altaffilmark{3}, 
Tae-Soo Pyo\altaffilmark{3}, 
Eugene Serabyn\altaffilmark{20}, 
Takuya Suenaga\altaffilmark{17}, 
Hiroshi Suto\altaffilmark{12}, 
Ryuji Suzuki\altaffilmark{21}, 
Yasuhiro Takahashi\altaffilmark{22}, 
Michihiro Takami\altaffilmark{23}, 
Naruhisa Takato\altaffilmark{3}, 
Hiroshi Terada\altaffilmark{3}, 
Daego Tomono\altaffilmark{3}, 
Edwin~L. Turner\altaffilmark{1,27}, 
Makoto Watanabe\altaffilmark{24}, 
John Wisniewski\altaffilmark{25}, 
Toru Yamada\altaffilmark{26}, 
Hideki Takami\altaffilmark{3}, 
Motohide Tamura\altaffilmark{12}
}

\altaffiltext{1}{Department of Astrophysical Sciences, Princeton University, NJ 08544, USA; \texttt{janson@astro.princeton.edu}}
\altaffiltext{2}{Department of Astrophysics, CAB (INTA-CSIC), Instituto Nacional de T\'ecnica Aerospacial, Torrej\'onde Ardoz, 28850, Madrid, Spain}
\altaffiltext{3}{Subaru Telescope, 650 North Aohoku Place, Hilo, HI 96720, USA}
\altaffiltext{4}{Astronomical Institute ``Anton Pannekoek'', University of Amsterdam, Science Park 904, 1098 XH Amsterdam, The Netherlands}
\altaffiltext{5}{Department of Physics and Astronomy, College of Charleston, 58 Coming Street, Charleston, SC 29424, USA}
\altaffiltext{6}{Universit\"ats-Sternwarte M\"unchen, Ludwig-Maximilians-Universit\"at, Scheinerstr. 1, 81679 Munich, Germany}
\altaffiltext{7}{Department of Astronomy and Astrophysics, University of Toronto, 50 St. George St, M5S 3H4, Toronto ON, Canada}
\altaffiltext{8}{Exoplanets and Stellar Astrophysics Laboratory, Code 667, Goddard Space Flight Center, Greenbelt, MD 2071, USA}
\altaffiltext{9}{Nishi-Harima Astronomical Observatory, Center for Astronomy, University of Hyogo, 407-2 Nishigaichi, Sayo, Hyogo 679-5313, Japan}
\altaffiltext{10}{Department of Earth and Space Science, Graduate School of Science, Osaka University, 1-1 Machikaneyama, Toyonaka, Osaka 560-0043, Japan}
\altaffiltext{11}{Department of Physics, University of Notre Dame, 225 Nieuwland Science Hall, Notre Dame, IN 46556, USA}
\altaffiltext{12}{National Astronomical Observatory of Japan, 2-21-1 Osawa, Mitaka, Tokyo 181-8588, Japan}
\altaffiltext{13}{Laboratoire Lagrange, UMR7239, University of Nice-Sophia Antipolis, CNRS, Observatoire de la Cote d'Azur, 06300 Nice, France}
\altaffiltext{14}{Max Planck Institute for Astronomy, K\"onigstuhl 17, 69117 Heidelberg, Germany}
\altaffiltext{15}{Eureka Scientific, 2452 Delmer, Suite 100, Oakland CA 96002, USA}
\altaffiltext{16}{Institute for Astronomy, University of Hawai`i, 640 North A'ohoku Place, Hilo, HI 96720, USA}
\altaffiltext{17}{Department of Astronomical Science, Graduate University for Advanced Studies (Sokendai), Tokyo 181-8588, Japan}
\altaffiltext{18}{Department of Astronomy, Kyoto University, Kitsahirakawa-Oiwake-cho, Sakyo-ku, Kyoto, 606-8502, Japan}
\altaffiltext{19}{Office of the President, Hiroshima University, 1-3-2 Kagamiyama, Hagashi-Hiroshima, 739-8511, Japan}
\altaffiltext{20}{Jet Propulsion Laboratory, California Institute of Technology, Pasadena, CA 91109, USA}
\altaffiltext{21}{TMT Observatory Corporation, 1111 South Arroyo Parkway, Pasadena, CA 91105, USA}
\altaffiltext{22}{Department of Astronomy, The University of Tokyo, Hongo 7-3-1, Bunkyo-ku, Tokyo 113-0033, Japan}
\altaffiltext{23}{Institute of Astronomy and Astrophysics, Academia Sinica, P.O. Box 23-141, Taipei 106, Taiwan}
\altaffiltext{24}{Department of Cosmosciences, Hokkaido University, Sapporo 060-0810, Japan}
\altaffiltext{25}{H.L. Dodge Department of Physics and Astronomy, University of Oklahoma, 440 W Brooks St Norman, OK 73019, USA}
\altaffiltext{26}{Astronomical Institute, Tohoku University, Aoba, Sendai 980-8578, Japan}
\altaffiltext{27}{Kavli Institute for the Physics and Mathematics of the Universe, The University of Tokyo, Kashiwa 277-8568, Japan}
\altaffiltext{28}{Hubble fellow}

\begin{abstract}\noindent
Debris disks around young main-sequence stars often have gaps and cavities which for a long time have been interpreted as possibly being caused by planets. In recent years, several giant planet discoveries have been made in systems hosting disks of precisely this nature, further implying that interactions with planets could be a common cause of such disk structures. As part of the SEEDS high-contrast imaging survey, we are surveying a population of debris disk-hosting stars with gaps and cavities implied by their spectral energy distributions, in order to attempt to spatially resolve the disk as well as to detect any planets that may be responsible for the disk structure. Here we report on intermediate results from this survey. Five debris disks have been spatially resolved, and a number of faint point sources have been discovered, most of which have been tested for common proper motion, which in each case has excluded physical companionship with the target stars. From the detection limits of the 50 targets that have been observed, we find that $\beta$~Pic~b-like planets ($\sim$10~$M_{\rm jup}$ planets around G--A-type stars) near the gap edges are less frequent than 15--30\%, implying that if giant planets are the dominant cause of these wide (27~AU on average) gaps, they are generally less massive than $\beta$~Pic~b.
\end{abstract}


\section{Introduction}
\label{s:intro}

The close circumstellar environment around mature (post-T~Tauri and Herbig Ae/Be) stars has traditionally been difficult to study directly, due to the strong flux from the star itself, which drowns out the light of its physical surroundings over a wide range of wavelengths. However, developments in high-contrast and high-resolution instruments and techniques have made this environment increasingly accessible to detailed study in recent years. Several surveys have been performed \citep[e.g.][]{kasper2007,lafreniere2007b,rameau2013} and a number of extrasolar planets have been imaged by now \citep[e.g.][]{marois2010,lagrange2010,carson2013}, and while the most extreme debris disk systems have been possible to image for some time \citep[e.g.][]{smith1984}, the sample of spatially resolved disks is presently growing rapidly, both in thermal \citep[e.g.][]{greaves2005,wilner2011,acke2012} and scattered radiation \citep[e.g.][]{krist2005,kalas2005,buenzli2010}. Nonetheless, most planets and disks are still discovered only indirectly, through stellar radial velocity or transits in the case of planets \citep[e.g.][]{mayor1995,borucki2011} and through infrared excess in the case of disks \citep[e.g.][]{beichman2006,su2006}.

The Strategic Exploration of Exoplanets and Disks with Subaru \citep[SEEDS;][]{tamura2009} is a large-scale survey using adaptive optics (AO) assisted high-contrast imaging for studying planets and disks, from primordial and transitional systems \citep[e.g.][]{kusakabe2012,muto2012,grady2013} to mature systems. A sub-survey of this larger effort concerns the study of debris disk systems. This study has several purposes, including: 1) searching for direct light from debris disks, in the sense of acquiring spatially resolved images of disks that have previously only been identified from infrared excess, 2) searching for planets in systems with known debris disks, and 3) studying interactions and correlations between planets and debris disks. Interestingly, many of the recently imaged planets coincide with debris disks \citep{marois2008,marois2010,lagrange2009}. Many disks also have morphological indications of the presence of dynamical influence from planets in the system \citep[e.g.][]{hines2007,buenzli2010,thalmann2011,currie2012b,quanz2013}, such as eccentric gaps with sharp inner boundaries or apparently resonant dust concentrations \citep[e.g.][]{quillen2002,quillen2006}, although alternative mechanisms have been suggested \citep[e.g.][]{jalali2012,lyra2012}. Thus, stars hosting debris disks are promising targets for imaging of massive exoplanets.

In previous publications, we have presented two results from the debris disk survey, in the form of spatially resolved disks around HR~4796~A \citep{thalmann2011} and HIP~79977 \citep{thalmann2013}. Here, we will summarize the results from the rest of the survey so far, including images of spatially resolved disks and detection limits for planets which are interpreted in the context of the disk architecture in the system, and which form part of the basis for a statistical study that is presently in progress (Brandt et al., in prep.). In the following, we first describe the target selection in Sect. \ref{s:targets} and the observations and data reduction in Sect. \ref{s:observations}, followed by a presentation of the results in Sect. \ref{s:results}. We discuss and summarize our results in Sect. \ref{s:discussion}.

\section{Target Selection}
\label{s:targets}

\begin{figure*}[p]
\centering
\includegraphics[width=17cm]{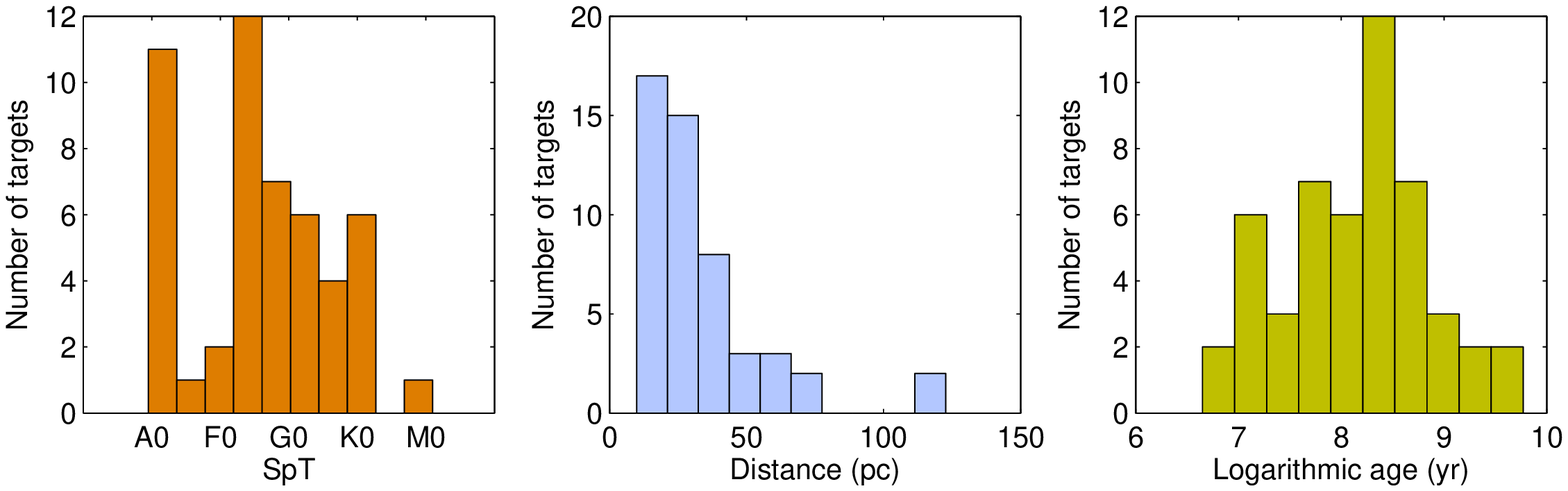}
\caption{Histograms showing the distributions of the sample in spectral type, distance, and age. The ages plotted here are the geometrical means of the lower and upper age limits derived for each target.}
\label{f:hist_spt}
\end{figure*}

A master list of targets was compiled from a wide range of literature sources identifying debris disk host stars based on infrared excess as measured by telescopes such as IRAS and Spitzer \citep[e.g.][]{rieke2005,rhee2007,trilling2008,plavchan2009}. Targets for specific SEEDS runs were then selected continuously from this list, prioritized on the basis of disk properties (fractional luminosity and predicted angular separation) and possible planet properties (ease of detection, based on proximity and youth, as well as stellar mass assuming a constant typical planet-star mass ratio). Special emphasis was placed on cold disks, characterized by the presence of dust at large physical separations but indications of gaps or cavities at smaller separations. Such gaps could be caused by planets \citep[see e.g.][and references therein.]{apai2008} which could in turn be observable in high-contrast images. A few warm disks however were also observed -- these could have planets at larger separations, and the disk in such systems should be highly luminous at small separations, where HiCIAO performs the most competitively. Some high-profile planet-search targets were purposefully omitted -- these are cases where specialized deep observations have been performed in dedicated studies, upon which it would be difficult or impossible to improve in a general survey with a 1~h observation in $H$-band. In particular, this is true for the targets $\epsilon$~Eri and Vega \citep{janson2008,heinze2008}. The special case of Fomalhaut \citep[e.g.][]{kalas2008,janson2012,currie2012a,galicher2013,kenworthy2013} was also omitted for this reason. Histograms for the spectral type, distance, and age of the targets are shown in Fig. \ref{f:hist_spt}.

\section{Observations and Data Reduction}
\label{s:observations}

The observations were carried out as part of the SEEDS program at the Subaru Telescope, using the HiCIAO camera \citep{tamura2006,hodapp2008} with the AO188 adaptive optics system \citep{hayano2008}. The bulk of observations were taken throughout 2011 and 2012, with some observations also taken in 2009 and 2010 (see Table \ref{t:observations}). No mask was used, but the detector was instead allowed to saturate at the PSF core, typically out to a radius of 0.3\arcsec. All observations made use of the angular differential imaging technique \citep[ADI;][]{marois2006} with the pupil fixed on the detector, and were performed using the $H$-band filter, with a central wavelength of 1.65~$\mu$m and a bandwidth of 0.29~$\mu$m. In most cases, the instrument was set to direct imaging, but in a few cases, the polarimetric differential imaging (PDI) mode was used, in which the beam is split into two orthogonal polarization states using a Wollaston prism, with each corresponding image mapped onto one half of the detector. In those cases, the results presented here are based on separate reductions of each polarization state, which we then average together. We do not include any PDI reductions in this study. The typical telescope time spent on a target was $\sim$1~hour including overheads.

\begin{table*}[p]
\caption{Observing log.}
\label{t:observations}
\centering
\small{
\begin{tabular}{lcccccccc}
\hline
\hline
HD ID & HIP ID & Alt ID & R.A. & Dec & $N_{\rm f}$\tablenotemark{a} & $t_{\rm tot}$\tablenotemark{a} & Rot\tablenotemark{a} & Date \\
 & & & (hh mm ss) & (dd mm ss) & & (min) & (deg) & \\
\hline
HD 377	&	HIP 682	&	---	&	00 08 25.7455	&	+06 37 00.498	&	31	&	10.3	&	9.28	&	2010-12-02	\\
HD 7590	&	HIP 5944	&	V445 And	&	01 16 29.2530	&	+42 56 21.911	&	87	&	11.6	&	33.8	&	2011-09-04	\\
HD 8907	&	HIP 6878	&	---	&	01 28 34.3597	&	+42 16 03.677	&	86	&	30.1	&	33.6	&	2012-01-02	\\
HD 9672	&	HIP 7345	&	49 Cet	&	01 34 37.7788	&	-15 40 34.893	&	195	&	32.5	&	26.2	&	2011-12-24	\\
HD 10008	&	HIP 7576	&	EX Cet	&	01 37 35.4661	&	-06 45 37.525	&	65	&	10.8	&	32.8	&	2010-12-02	\\
HD 12039	&	HIP 9141	&	DK Cet	&	01 57 48.9784	&	-21 54 05.345	&	270	&	45.0	&	21.5	&	2012-09-11	\\
HD 15115	&	HIP 11360	&	---	&	02 26 16.2447	&	+06 17 33.188	&	306	&	28.4	&	51.3	&	2009-12-25	\\
HD 15115	&	---	&	---	&	---	&	---	&	290	&	6.7	&	45.4	&	2009-12-25	\\
HD 15745	&	HIP 11847	&	---	&	02 32 55.8103	&	+37 20 01.045	&	57	&	19.0	&	25.7	&	2011-09-06	\\
HD 17925	&	HIP 13402	&	EP Eri	&	02 52 32.1287	&	-12 46 10.972	&	82	&	8.2	&	26.0	&	2011-09-06	\\
HD 25457	&	HIP 18859	&	HR 1249	&	04 02 36.745	&	-00 16 08.12	&	840	&	21.0	&	37.1	&	2012-09-13	\\
HD 281691	&	---	&	V1197 Tau	&	04 09 09.7402	&	+29 01 30.345	&	940	&	156.7	&	79.1	&	2012-11-07	\\
HD 31295	&	HIP 22845	&	7 Ori	&	04 54 53.7279	&	+10 09 02.999	&	213	&	5.3	&	27.5	&	2011-11-20	\\
HD 40136	&	HIP 28103	&	$\eta$ Lep	&	05 56 24.2930	&	-14 10 03.719	&	780	&	19.5	&	27.9	&	2012-11-05	\\
HD 60737	&	HIP 37170	&	---	&	07 38 16.4417	&	+47 44 55.230	&	71	&	17.8	&	19.3	&	2012-01-01	\\
HD 69830	&	HIP 40693	&	LHS 245	&	08 18 23.9473	&	-12 37 55.824	&	159	&	14.8	&	26.7	&	2010-01-23	\\
HD 70573	&	---	&	V748 Hya	&	08 22 49.951	&	+01 51 33.55	&	66	&	16.5	&	44.6	&	2011-01-30	\\
HD 73350	&	HIP 42333	&	V401 Hya	&	08 37 50.2932	&	-06 48 24.786	&	400	&	33.3	&	25.5	&	2011-12-30	\\
HD 73752	&	HIP 42430	&	LHS 5139A	&	08 39 07.9003	&	-22 39 42.750	&	231	&	9.6	&	12.0	&	2011-03-25	\\
HD 72905	&	HIP 42438	&	3 Uma	&	08 39 11.7040	&	+65 01 15.264	&	7742	&	193.6	&	17.3	&	2011-12-24	\\
HD 76151	&	HIP 43726	&	NLTT 20504	&	08 54 17.9475	&	-05 26 04.054	&	640	&	16.0	&	27.9	&	2011-12-26	\\
HD 88215	&	HIP 49809	&	HR 3991	&	10 10 05.8864	&	-12 48 57.324	&	510	&	21.3	&	21.3	&	2011-12-31	\\
HD 91312	&	HIP 51658	&	HR 4132	&	10 33 13.8883	&	+40 25 32.016	&	750	&	18.8	&	35.0	&	2012-05-12	\\
HD 92945	&	HIP 52462	&	V419 Hya	&	10 43 28.2717	&	-29 03 51.421	&	79	&	13.2	&	16.5	&	2011-12-25	\\
HD 102647	&	HIP 57632	&	$\beta$ Leo	&	11 49 03.5776	&	+14 34 19.417	&	82	&	1.9	&	115.0	&	2010-01-24	\\
HD 104860	&	HIP 58876	&	---	&	12 04 33.7302	&	+66 20 11.720	&	64	&	21.3	&	20.8	&	2012-04-11	\\
HD 106591	&	HIP 59774	&	$\delta$ Uma	&	12 15 25.5601	&	+57 01 57.421	&	250	&	11.6	&	23.5	&	2010-01-25	\\
HD 106591	&	---	&	---	&	---	&	---	&	369	&	9.2	&	22.8	&	2011-01-28	\\
HD 107146	&	HIP 60074	&	NLTT 30317	&	12 19 06.5015	&	+16 32 53.869	&	160	&	37.1	&	34.3	&	2009-12-24	\\
HD 107146	&	---	&	---	&	---	&	---	&	246	&	20.5	&	122.3	&	2011-03-25	\\
HD 109085	&	HIP 61174	&	$\eta$ Crv	&	12 32 04.2270	&	-16 11 45.627	&	71	&	4.9	&	21.3	&	2010-01-23	\\
HD 109573	&	HIP 61498	&	HR 4796A	&	12 36 01.0316	&	-39 52 10.219	&	87	&	14.5	&	23.5	&	2011-05-24	\\
HD 110411	&	HIP 61960	&	$\rho$ Vir	&	12 41 53.0565	&	+10 14 08.251	&	183	&	7.6	&	63.0	&	2011-01-29	\\
HD 112429	&	HIP 63076	&	IR Dra	&	12 55 28.5486	&	+65 26 18.505	&	258	&	6.5	&	20.6	&	2011-05-24	\\
HD 113337	&	HIP 63584	&	HR 4934	&	13 01 46.9269	&	+63 36 36.810	&	159	&	13.3	&	19.0	&	2011-05-21	\\
HD 113337	&	---	&	---	&	---	&	---	&	174	&	14.5	&	20.0	&	2012-02-27	\\
HD 125162	&	HIP 69732 	&	NLTT 36818	&	14 16 23.0187	&	+46 05 17.900	&	225	&	7.5	&	28.6	&	2011-01-30	\\
HD 127821	&	HIP 70952	&	NLTT 37640	&	14 30 46.0702	&	+63 11 08.836	&	130	&	10.8	&	18.9	&	2011-05-26	\\
HD 128167	&	HIP 71284	&	$\sigma$ Boo	&	14 34 40.8171	&	+29 44 42.468	&	730	&	18.3	&	75.3	&	2012-04-11	\\
HD 128311	&	HIP 71395	&	HN Boo	&	14 36 00.5607	&	+09 44 47.466	&	180	&	15.0	&	66.2	&	2012-02-27	\\
HD 135599	&	HIP 74702	&	V379 Ser	&	15 15 59.1667	&	+00 47 46.905	&	198	&	13.9	&	42.6	&	2011-05-25	\\
HD 135599	&	---	&	---	&	---	&	---	&	231	&	19.3	&	54.6	&	2012-02-28	\\
HD 139006	&	HIP 76267	&	$\alpha$ CrB	&	15 34 41.2681	&	+26 42 52.895	&	460	&	11.5	&	83.5	&	2012-04-12	\\
HD 139664	&	HIP 76829	&	NLTT 40843	&	15 41 11.3774	&	-44 39 40.338	&	240	&	6.0	&	15.5	&	2011-05-22	\\
HD 141569	&	HIP 77542	&	---	&	15 49 57.7489	&	-03 55 16.360	&	74	&	12.3	&	33.2	&	2011-03-26	\\
HD 146897	&	HIP 79977	&	---	&	16 19 29.2425	&	-21 24 13.264	&	60	&	20.0	&	19.6	&	2012-05-12	\\
HD 146897	&	---	&	---	&	---	&	---	&	69	&	34.5	&	18.3	&	2012-07-07	\\
HD 152598	&	HIP 82587	&	53 Her	&	16 52 58.0578	&	+31 42 06.026	&	630	&	21.0	&	43.0	&	2012-05-11	\\
HD 161868	&	HIP 87108	&	$\gamma$ Oph	&	17 47 53.5605	&	+02 42 26.194	&	800	&	20.0	&	41.3	&	2012-07-11	\\
HD 162917	&	HIP 87558	&	HR 6670	&	17 53 14.1849	&	+06 06 05.127	&	243	&	17.0	&	57.4	&	2012-07-09	\\
HD 175742	&	HIP 92919 	&	V775 Her	&	18 55 53.2247	&	+23 33 23.940	&	87	&	14.5	&	104.3	&	2011-05-23	\\
HD 175742	&	---	&	---	&	---	&	---	&	222	&	37.0	&	123.8	&	2012-05-11	\\
HD 183324	&	HIP 95793	&	V1431 Aql	&	19 29 00.9882	&	+01 57 01.611	&	276	&	36.8	&	34.7	&	2012-07-10	\\
HD 192263	&	HIP 99711	&	V1703 Aql	&	20 13 59.846	&	-00 52 00.75	&	77	&	12.8	&	44.1	&	2012-05-14	\\
HD 197481	&	HIP 102409	&	AU Mic	&	20 45 09.5318	&	-31 20 27.238	&	53	&	25.8	&	11.7	&	2009-11-01	\\
HD 206860	&	HIP 107350	&	HN Peg	&	21 44 31.3299	&	+14 46 18.981	&	207	&	8.6	&	72.1	&	2011-08-03	\\
HD 207129	&	HIP 107649	&	NLTT 52100	&	21 48 15.7514	&	-47 18 13.014	&	232	&	5.8	&	16.6	&	2011-08-02	\\
\hline
\end{tabular}
}
\tablenotetext{a}{\scriptsize{$N_{\rm f}$ denotes the number of frames used, $t_{\rm tot}$ the total integration time, and $\theta _{\rm r}$ the field rotation angle during the observation.}}
\end{table*}

\begin{figure}[p]
\centering
\includegraphics[width=8cm]{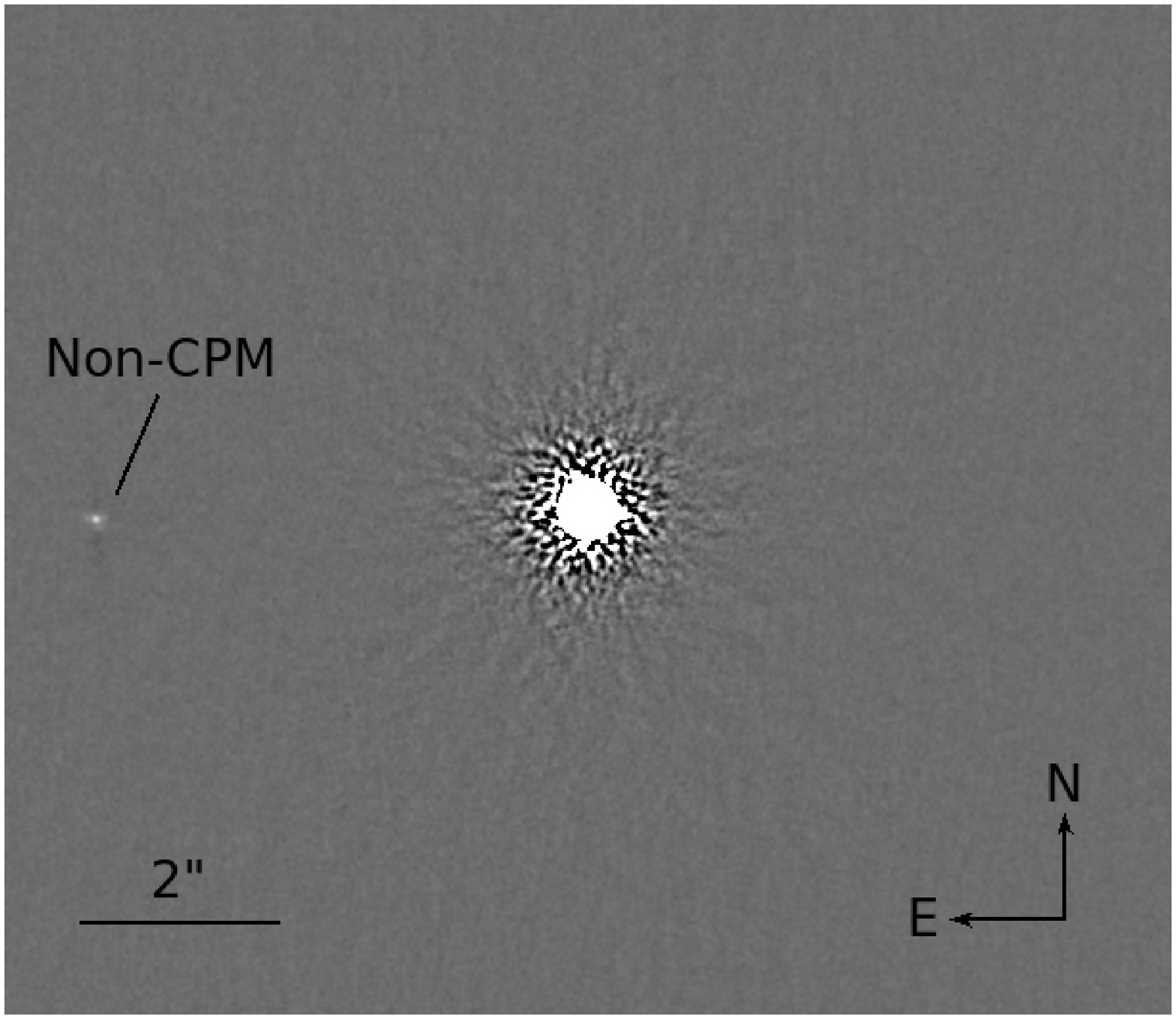}
\caption{Example of a final reduced image using the ACORNS-ADI pipeline, showing the residual PSF noise of the star HD~113337 and a faint point source to the east. The point source does not share a common proper motion with the primary (abbreviated as non-CPM), hence it is a physically unrelated field object (see Fig. \ref{f:hr4934astro}).}
\label{f:hr4934im}
\end{figure}

The ADI reductions were uniformly performed using the ACORNS-ADI pipeline \citep{brandt2013}, with the same procedure as given in the ACORNS paper. As a brief summary, the data were destriped\footnote{Removal of correlated read-noise, which causes striping in the images.}, flat fielded and corrected for field distortion. Relative centroiding was done using PSF fitting on non-saturated parts of the PSF, and absolute centering was based on visual inspection with a $\sim$0.5~pixel precision. PSF subtraction was performed with a LOCI-based scheme \citep[Locally Optimized Combination of Images;][]{lafreniere2007a}. As LOCI parameters we used a PSF FWHM of 6 pixels, an angular protection zone of 0.7 FWHM, and 200 PSF footprint optimization regions. Individual PSF-subtracted frames were de-rotated and combined using a trimmed mean approach to produce the final image (see Fig. \ref{f:hr4934im} for an example). For each final image of a given target, an S/N-map was produced by dividing the signal at all positions by the local noise (calculated in an annulus at the corresponding separation). In this process, we include a correction for the signal attenuation imposed by the LOCI algorithm. The S/N-map provides a data format in which point sources can be easily identified and in which it can be determined whether or not they are statistically significant. Detection limits for a 5.5$\sigma$ criterion were produced by normalizing the radial noise profiles by the primary brightness, which was determined from non-saturated exposures acquired before and after each ADI sequence.

In cases where candidates were present in the HiCIAO images and the targets had been previously observed with AO-assisted imagers, we analyzed the archival images using similar procedures as above, but adapted to the respective telescopes and instruments. Images from Gemini/NIRI \citep{hodapp2003}, Gemini/NICI \citep[e.g.][]{artigau2008}, Keck/NIRC2 \citep[e.g.][]{mclean2003}, Subaru/IRCS \citep{kobayashi2000} and HST/NICMOS \citep[e.g.][]{schultz2003} were used for this purpose, saving several hours of Subaru telescope time that would otherwise have been necessary for executing follow-up observations in those cases, and thus demonstrating the broad utility of archiving data from large telescopes.

\section{Results}
\label{s:results}

\subsection{General Results}
\label{s:general}

\begin{figure}[p]
\centering
\includegraphics[width=8cm]{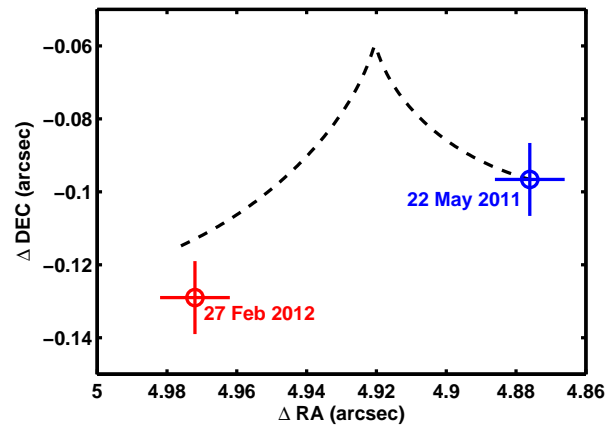}
\caption{Example of astrometric analysis, for the case of HD~113337. The second epoch observation falls close to the expected motion for a static background object (dashed line), and is clearly inconsistent with common proper motion. Thus, it can be concluded that the point source is physically unrelated to HD~113337.}
\label{f:hr4934astro}
\end{figure}

As can be generally expected in a survey of this kind, many faint point sources are detected in the images, the majority of which are physically unrelated background stars. Due to the fact that the contaminant fraction increases rapidly with angular separation from the parent star, small angular separations have been prioritized for follow-up. Companion candidates that were detected in the data inside of 5\arcsec\ and with a $>$5.5$\sigma$ significance were checked for common proper motion by either using archival data when available, or second epoch observations over a $\sim$1 year baseline (see Fig. \ref{f:hr4934astro} for an example). No substellar companions have been verified so far among the targets. One target remains for which candidates inside of the priority region have not yet been followed up; HD~162917 was observed in late 2012 and will be re-observed at a later stage. Given the low galactic latitude of this target, the candidates are likely to be background stars. The point sources are listed in Table \ref{t:pointsources}.

\begin{table*}[p]
\caption{Properties of the imaged point sources.}
\label{t:pointsources}
\centering
\small{
\begin{tabular}{lccccc}
\hline
\hline
HD ID & CC & $\Delta H$ (mag) & $\Delta$ R.A. (\arcsec) & $\Delta$ Dec. (\arcsec) & Epoch \\
\hline
HD 15745	&	1	&	11.1$\pm$0.1	&	-1.85$\pm$0.01	&	-0.63$\pm$0.01	&	2011-09-06	\\
HD 60737	&	1	&	10.3$\pm$0.1	&	6.29$\pm$0.01	&	-3.02$\pm$0.01	&	2012-01-01	\\
HD 69830	&	1	&	13.4$\pm$0.1	&	-5.73$\pm$0.01	&	-3.91$\pm$0.01	&	2010-01-23	\\
HD 70573	&	1	&	13.8$\pm$0.2	&	2.61$\pm$0.01	&	-2.24$\pm$0.01	&	2011-01-30	\\
HD 73350	&	1	&	11.7$\pm$0.1	&	2.90$\pm$0.01	&	5.23$\pm$0.01	&	2011-12-30	\\
HD 73752	&	1	&	1.2$\pm$0.1	&	0.67$\pm$0.01	&	0.80$\pm$0.01	&	2011-03-25	\\
HD 73752	&	2	&	13.7$\pm$0.1	&	-4.50$\pm$0.01	&	6.02$\pm$0.01	&	2011-03-25	\\
HD 88215	&	1	&	14.5$\pm$0.1	&	-7.47$\pm$0.01	&	-0.89$\pm$0.01	&	2011-12-31	\\
HD 104860	&	1	&	12.1$\pm$0.1	&	-3.10$\pm$0.01	&	-0.55$\pm$0.01	&	2012-04-11	\\
HD 106591	&	1	&	15.1$\pm$0.1	&	3.22$\pm$0.01	&	-1.25$\pm$0.01	&	2010-01-25	\\
HD 106591	&	1	&	15.1$\pm$0.1	&	3.08$\pm$0.01	&	-1.30$\pm$0.01	&	2011-01-28	\\
HD 106591	&	2	&	15.9$\pm$0.1	&	1.26$\pm$0.01	&	-5.57$\pm$0.01	&	2010-01-25	\\
HD 106591	&	2	&	16.1$\pm$0.2	&	1.06$\pm$0.01	&	-5.59$\pm$0.01	&	2011-01-28	\\
HD 107146	&	1	&	14.9$\pm$0.1	&	-3.69$\pm$0.01	&	-5.07$\pm$0.01	&	2011-03-25	\\
HD 113337	&	1	&	13.6$\pm$0.1	&	4.88$\pm$0.01	&	-0.10$\pm$0.01	&	2011-05-21	\\
HD 113337	&	1	&	13.4$\pm$0.1	&	4.97$\pm$0.01	&	-0.13$\pm$0.01	&	2012-02-27	\\
HD 128311	&	1	&	12.4$\pm$0.1	&	4.33$\pm$0.01	&	-6.38$\pm$0.01	&	2012-02-27	\\
HD 141569	&	1	&	2.2$\pm$0.1	&	-5.50$\pm$0.01	&	5.23$\pm$0.01	&	2011-03-26	\\
HD 161868	&	1	&	14.2$\pm$0.1	&	-6.10$\pm$0.01	&	-0.10$\pm$0.01	&	2012-07-11	\\
HD 161868	&	2	&	14.8$\pm$0.1	&	6.05$\pm$0.01	&	3.89$\pm$0.01	&	2012-07-11	\\
HD 162917	&	1	&	12.0$\pm$0.1	&	2.46$\pm$0.01	&	-1.67$\pm$0.01	&	2012-07-09	\\
HD 162917	&	2	&	12.5$\pm$0.1	&	-2.73$\pm$0.01	&	2.06$\pm$0.01	&	2012-07-09	\\
HD 162917	&	3	&	12.8$\pm$0.1	&	0.35$\pm$0.01	&	-4.41$\pm$0.01	&	2012-07-09	\\
HD 175742	&	1	&	10.6$\pm$0.1	&	1.72$\pm$0.01	&	1.97$\pm$0.01	&	2011-05-23	\\
HD 175742	&	1	&	10.8$\pm$0.1	&	1.59$\pm$0.01	&	2.24$\pm$0.01	&	2012-05-11	\\
HD 183324	&	1	&	13.7$\pm$0.1	&	-0.73$\pm$0.01	&	1.71$\pm$0.01	&	2012-07-10	\\
HD 183324	&	2	&	14.6$\pm$0.1	&	3.25$\pm$0.01	&	1.36$\pm$0.01	&	2012-07-10	\\
HD 183324	&	3	&	14.6$\pm$0.1	&	3.40$\pm$0.01	&	-1.16$\pm$0.01	&	2012-07-10	\\
HD 183324	&	4	&	13.6$\pm$0.1	&	1.49$\pm$0.01	&	-4.29$\pm$0.01	&	2012-07-10	\\
HD 183324	&	5	&	13.9$\pm$0.1	&	-4.44$\pm$0.01	&	-3.15$\pm$0.01	&	2012-07-10	\\
HD 183324	&	6	&	15.2$\pm$0.2	&	5.00$\pm$0.01	&	4.43$\pm$0.01	&	2012-07-10	\\
HD 192263	&	1	&	13.6$\pm$0.1	&	-4.41$\pm$0.01	&	-5.83$\pm$0.01	&	2012-05-14	\\
HD 206860	&	1	&	15.3$\pm$0.2	&	1.69$\pm$0.01	&	2.45$\pm$0.01	&	2011-08-03	\\
HD 281691	&	1	&	1.7$\pm$0.1	&	4.33$\pm$0.01	&	5.22$\pm$0.01	&	2012-11-07	\\
\hline
\end{tabular}
}
\end{table*}

In some cases, the debris disk itself could be spatially resolved in our images. Two of these detections have been analysed in particular detail and published separately: HR~4796~A \citep{thalmann2011} and HIP~79977 \citep{thalmann2013}. Three other targets for which secure disk detections could be made are HD~15115, AU~Mic, and HD~141569. These cases are discussed in the individual notes below. The disk detection space of our survey has a very good complementarity to that of the Hubble Space Telescope (HST). HST is able to observe at visible wavelengths with exquisite sensitivity and has a PSF which is unaffected by the atmosphere, which means that it can observe faint and smooth disk emission. Such emission is much more difficult to observe from the ground, since our near-infrared observations are more sensitivity limited, in addition to the fact that PSF variations due to varying seeing are very similar in their characteristics to smooth disk material. In addition, this study has made use of ADI, which benefits the detection of sharp features in the disk while strongly self-subtracting smooth emission, particularly if it is azimuthally symmetric. On the other hand, the high contrast and spatial resolution of HiCIAO allows for detection of disks and disk features at small angular separations, where HST is unable to provide a comparable performance. Hence, we are unable to detect large-scale, smooth, and low-inclination structures such as the second ring of the HD~141569 disk, but can provide novel results on small-scale, sharp and high-inclination features such as the inner region of the HIP~79977 disk.

\begin{figure}[p]
\centering
\includegraphics[width=8cm]{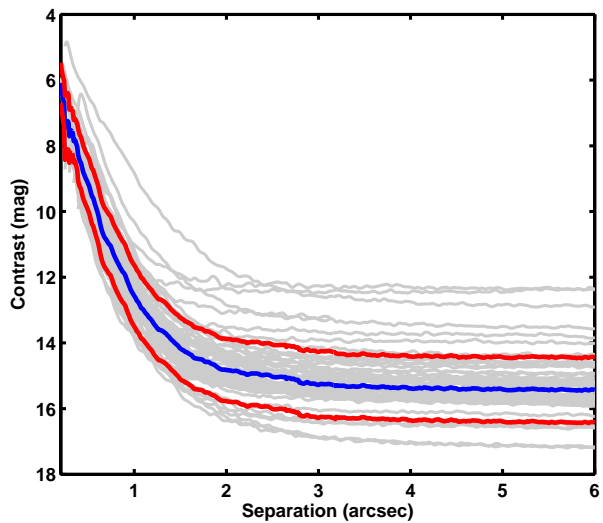}
\caption{Contrast as a function of angular separation. The individual contrast curves are shown in light gray. The thick blue line denotes the median contrast, and the red lines are separated from the median by one standard deviation of the curve-to-curve scatter in each direction. The target LHS~5139, where the binary companion affects the azimuthally averaged contrast to a signficiant extent, has been omitted from the figure.}
\label{f:dd_ccurves}
\end{figure}

The contrast performances for point sources are shown in Fig. \ref{f:dd_ccurves} and listed in Table \ref{t:contrasts}. The achievable contrast is largely dependent on the field rotation during an observation. This is due to the connection between field rotation and ADI perfomance. An increased total field rotation benefits ADI because it maximizes the number of reference frames in which the planet signature is sufficiently separated from its location in the target frame to be useful, which helps as long as the number of reference frames does not become so large that the LOCI optimization becomes over-constrained. In our reductions, we avoid this over-constraining by limiting the number of reference frames for each target frame to $\sim$80, uniformly spread across the observing sequence, which we find to produce roughly optimal performance. The contrast depends not only on the total field rotation, but also on the rotation rate. This is caused by the fact that frames taken over a small time span tend to correlate better than frames taken over larger time spans. A larger rotation rate thus allows for usage of reference frames that are better correlated with the target frame. The rotation rate that can be acquired depends on the declination of the target -- a minimal $| \delta - l |$ provides a maximal rotation rate, where $\delta$ is the declination and $l$ the latitude of the telescope. 

\begin{table*}[p]
\caption{Contrast at a range of angular separations.}
\label{t:contrasts}
\centering
\small{
\begin{tabular}{lccccccccc}
\hline
\hline
HD ID & Ep. & 0.25\arcsec & 0.5\arcsec & 0.75\arcsec & 1.0\arcsec & 1.5\arcsec & 2.0\arcsec & 3.0\arcsec & 5.0\arcsec \\
 & & (mag) & (mag) & (mag) & (mag) & (mag) & (mag) & (mag) & (mag) \\
\hline
HD 377	&	1	&	---	&	---	&	9.6	&	11.3	&	12.9	&	13.9	&	14.1	&	14.3	\\
HD 7590	&	1	&	---	&	8.8	&	10.7	&	12.4	&	14.0	&	14.6	&	15.0	&	15.2	\\
HD 8907	&	1	&	7.8	&	9.5	&	11.5	&	13.1	&	14.7	&	15.3	&	15.6	&	15.7	\\
HD 9672	&	1	&	---	&	8.7	&	10.3	&	11.6	&	13.2	&	13.9	&	14.4	&	14.6	\\
HD 10008	&	1	&	7.0	&	8.8	&	10.6	&	12.1	&	13.8	&	14.4	&	14.7	&	14.9	\\
HD 12039	&	1	&	---	&	8.7	&	10.7	&	12.1	&	13.6	&	14.1	&	14.4	&	14.6	\\
HD 15115	&	1	&	5.8	&	7.8	&	9.6	&	11.0	&	11.9	&	12.2	&	12.3	&	12.4	\\
HD 15115	&	2	&	7.1	&	9.3	&	11.4	&	12.9	&	14.0	&	14.3	&	---	&	---	\\
HD 15745	&	1	&	7.6	&	9.2	&	11.2	&	12.7	&	14.1	&	14.5	&	14.9	&	15.0	\\
HD 17925	&	1	&	---	&	8.8	&	10.8	&	12.3	&	14.2	&	15.1	&	15.6	&	15.8	\\
HD 25457	&	1	&	---	&	9.4	&	11.2	&	12.9	&	14.3	&	14.8	&	15.3	&	15.4	\\
HD 281691	&	1	&	7.4	&	9.7	&	11.0	&	11.9	&	12.2	&	12.4	&	12.5	&	12.5	\\
HD 31295	&	1	&	5.0	&	6.5	&	7.7	&	8.8	&	10.6	&	11.8	&	12.6	&	12.8	\\
HD 40136	&	1	&	---	&	10.4	&	12.2	&	13.7	&	15.4	&	16.1	&	16.9	&	17.1	\\
HD 60737	&	1	&	---	&	8.7	&	10.8	&	12.3	&	13.9	&	14.4	&	14.8	&	14.9	\\
HD 69830	&	1	&	---	&	8.7	&	10.5	&	11.9	&	13.7	&	14.6	&	14.9	&	15.1	\\
HD 70573	&	1	&	6.5	&	8.6	&	10.6	&	12.0	&	13.3	&	13.6	&	13.9	&	14.0	\\
HD 73350	&	1	&	7.7	&	9.0	&	10.7	&	12.1	&	13.7	&	14.6	&	15.1	&	15.3	\\
HD 73752	&	1	&	---	&	6.4	&	6.9	&	6.8	&	10.4	&	12.5	&	13.8	&	14.2	\\
HD 72905	&	1	&	---	&	9.0	&	10.9	&	12.4	&	14.1	&	14.7	&	15.1	&	15.2	\\
HD 76151	&	1	&	---	&	8.9	&	10.6	&	12.0	&	13.6	&	14.3	&	14.6	&	14.7	\\
HD 88215	&	1	&	---	&	8.8	&	11.2	&	12.7	&	14.4	&	15.2	&	15.6	&	15.8	\\
HD 91312	&	1	&	---	&	9.3	&	11.3	&	12.8	&	14.6	&	15.3	&	15.6	&	15.7	\\
HD 92945	&	1	&	---	&	7.2	&	9.0	&	10.3	&	12.0	&	12.9	&	13.3	&	13.5	\\
HD 102647	&	1	&	---	&	---	&	11.8	&	13.6	&	15.4	&	16.3	&	16.9	&	17.1	\\
HD 104860	&	1	&	---	&	10.3	&	12.2	&	13.6	&	14.8	&	15.2	&	15.4	&	15.7	\\
HD 106591	&	1	&	---	&	8.9	&	11.5	&	13.1	&	15.1	&	16.0	&	16.4	&	16.5	\\
HD 106591	&	2	&	---	&	9.5	&	11.4	&	12.9	&	15.1	&	15.8	&	16.3	&	16.5	\\
HD 107146	&	1	&	6.9	&	8.8	&	10.8	&	12.4	&	13.8	&	14.4	&	---	&	---	\\
HD 107146	&	2	&	7.5	&	9.1	&	11.0	&	12.5	&	14.2	&	14.8	&	15.0	&	15.1	\\
HD 109085	&	1	&	---	&	---	&	10.8	&	12.3	&	14.1	&	15.0	&	15.4	&	15.8	\\
HD 109573	&	1	&	---	&	9.8	&	11.5	&	12.4	&	14.6	&	15.2	&	15.6	&	15.7	\\
HD 110411	&	1	&	---	&	8.8	&	10.7	&	12.1	&	14.2	&	15.0	&	15.4	&	15.5	\\
HD 112429	&	1	&	---	&	9.4	&	11.3	&	13.0	&	14.6	&	15.1	&	15.4	&	15.6	\\
HD 113337	&	1	&	---	&	9.6	&	11.4	&	12.9	&	14.5	&	15.2	&	15.5	&	15.7	\\
HD 113337	&	2	&	---	&	9.8	&	11.4	&	12.9	&	14.5	&	15.2	&	15.7	&	15.7	\\
HD 125162	&	1	&	---	&	7.9	&	9.6	&	11.1	&	13.3	&	14.3	&	15.0	&	15.2	\\
HD 127821	&	1	&	---	&	8.6	&	10.0	&	11.4	&	13.2	&	14.0	&	14.5	&	14.6	\\
HD 128167	&	1	&	---	&	10.1	&	12.0	&	13.6	&	15.2	&	16.0	&	16.4	&	16.5	\\
HD 128311	&	1	&	---	&	9.5	&	11.4	&	12.7	&	14.4	&	15.0	&	15.5	&	15.6	\\
HD 135599	&	1	&	8.4	&	10.7	&	12.5	&	13.6	&	14.9	&	15.3	&	15.6	&	15.6	\\
HD 135599	&	2	&	---	&	10.4	&	12.6	&	13.9	&	15.3	&	15.7	&	16.1	&	16.2	\\
HD 139006	&	1	&	---	&	9.9	&	12.1	&	13.7	&	15.6	&	16.4	&	16.9	&	---	\\
HD 139664	&	1	&	---	&	8.4	&	10.0	&	11.5	&	13.4	&	14.4	&	15.2	&	15.5	\\
HD 141569	&	1	&	8.7	&	9.9	&	11.9	&	13.0	&	14.0	&	14.4	&	14.6	&	14.7	\\
HD 146897	&	1	&	---	&	9.4	&	11.2	&	12.2	&	13.2	&	13.5	&	13.7	&	13.8	\\
HD 146897	&	2	&	---	&	8.4	&	10.5	&	11.6	&	12.7	&	13.1	&	13.2	&	---	\\
HD 152598	&	1	&	---	&	9.6	&	11.7	&	13.3	&	14.6	&	15.0	&	15.3	&	15.4	\\
HD 161868	&	1	&	---	&	9.1	&	11.0	&	12.6	&	14.2	&	15.0	&	15.4	&	15.6	\\
HD 162917	&	1	&	---	&	8.8	&	10.7	&	12.1	&	13.8	&	14.5	&	14.8	&	15.1	\\
HD 175742	&	1	&	7.8	&	10.2	&	12.1	&	13.5	&	14.6	&	15.0	&	15.3	&	15.4	\\
HD 175742	&	2	&	---	&	10.3	&	12.4	&	13.8	&	14.9	&	15.3	&	15.5	&	15.7	\\
HD 183324	&	1	&	---	&	9.7	&	11.6	&	13.0	&	14.3	&	14.8	&	15.2	&	15.3	\\
HD 192263	&	1	&	---	&	10.2	&	12.1	&	13.4	&	14.7	&	15.1	&	15.3	&	15.5	\\
HD 197481	&	1	&	---	&	8.5	&	11.1	&	12.5	&	14.2	&	14.8	&	15.5	&	---	\\
HD 206860	&	1	&	---	&	10.0	&	12.0	&	13.5	&	14.9	&	15.5	&	15.7	&	15.8	\\
HD 207129	&	1	&	---	&	9.2	&	11.1	&	12.6	&	14.2	&	14.9	&	15.3	&	15.4	\\
\hline
\end{tabular}
}
\end{table*}

\subsection{Individual Targets}

Below, we list individual notes concerning the results on different targets in the survey, such as detections of disks or point sources, as well as other details from the scientific literature that are relevant for the context.

\begin{figure}[p]
\centering
\includegraphics[width=8cm]{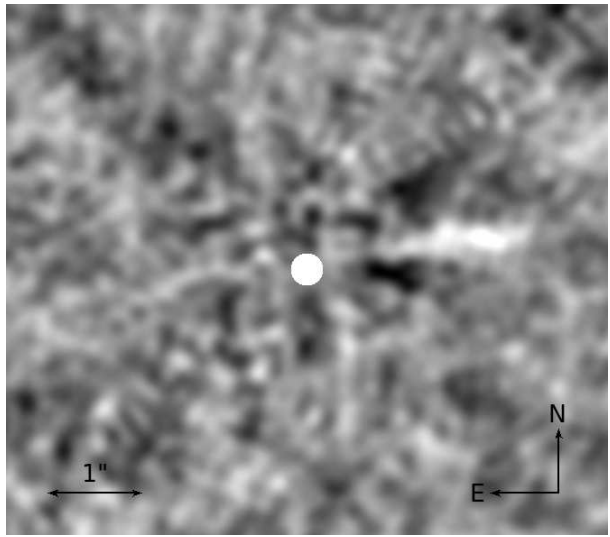}
\caption{Image of the disk around HD~15115. The $S/N$ is limited, but disk emission is seen at the expected region of maximal disk flux from previous images \citep[compare e.g.][]{rodigas2012}, on the western side of the star. A Gaussian smoothing kernel of 15 pixel FWHM has been applied to the data.}
\label{f:hd15115disk}
\end{figure}

\textbf{HD~15115 (HIP~11360)}: This star has a known debris disk which has been spatially resolved at several near-infrared wavelengths \citep[e.g.][]{kalas2007a,debes2008,rodigas2012}. We also detect the disk in the HiCIAO data (see Fig. \ref{f:hd15115disk}), but at limited $S/N$ which does not improve on the results in previous studies.

\textbf{HD~15745 (HIP~11847)}: The debris disk around HD~15745 has been spatially resolved in HST observations \citep{kalas2007b}, but is not visible in the HiCIAO images due to its smooth and azimuthally extended features. A candidate companion was seen at $\Delta \alpha$ = -1.85\arcsec\ and $\Delta \delta$ = -0.63\arcsec\ with HiCIAO. The point source is faint but visible in the archival HST images from 2004, where it has $\Delta \alpha$ = -1.65\arcsec\ and $\Delta \delta$ = -0.89\arcsec, demonstrating that it is a background contaminant. There is also an intermediate epoch available from Keck in 2007, with the point source located at $\Delta \alpha$ = -1.72\arcsec\ and $\Delta \delta$ = -0.75\arcsec, further confirming this conclusion.

\textbf{HD~60737 (HIP~37170)}: The field of HD~60737 is empty except for a point source at $\Delta \alpha$ = 6.29\arcsec\ and $\Delta \delta$ = -3.02\arcsec, which has already been identified as a background star by \citet{metchev2009}.

\textbf{HD~69830 (HIP~40693, LHS~245)}: This system is notable for its planetary system which contains three known planets so far, all with Neptune-like masses \citep{lovis2006}. It also hosts a warm debris disk \citep{beichman2006}, which has been resolved with interferometry \citep{smith2009}. Our images do not reveal the disk, and due to the probably quite old age of the system \citep[approximately 6~Gyr;][]{mamajek2008} and small physical scale ($\sim$1--2~AU) of the dust location, no stringent constraints can be drawn regarding planets near the disk edge from the imaging. We do detect a point source at $\Delta \alpha$ = -5.73\arcsec\ and $\Delta \delta$ = -3.91\arcsec. This candidate is visible in an archival HST image from 2007 with $\Delta \alpha$ = -5.24\arcsec\ and $\Delta \delta$ = -5.94\arcsec, hence it is a physically unrelated background star.

\textbf{HD~70573 (V748~Hya)}: There is an object at $\Delta \alpha$ = 2.61\arcsec\ and $\Delta \delta$ = -2.24\arcsec\ in the HiCIAO images. Although the source appears somewhat extended, we nonetheless examined archival data to test its nature. This turned up the object in archival NICI images, where it is located at $\Delta \alpha$ = 2.47\arcsec\ and $\Delta \delta$ = -2.38\arcsec, indeed implying non-common proper motion. This star has a planet candidate from radial velocity measurements, at a semi-major axis of 1.8~AU \citep{setiawan2007}.

\textbf{HD~73350 (HIP~42333, V401~Hya)}: There is a point source at $\Delta \alpha$ = 2.90\arcsec\ and $\Delta \delta$ = 5.23\arcsec\ in the HiCIAO data. It is considered of low priority due to its relatively large separation from the primary.

\textbf{HD~73752 (HIP~42430, LHS~5139)}: A known binary \citep[e.g.][]{mason2001}, the location of the secondary relative to the primary in the HiCIAO images is $\Delta \alpha$ = 0.67\arcsec\ and $\Delta \delta$ = 0.80\arcsec. There is another possible candidate in the image at $\Delta \alpha$ = -4.50\arcsec\ and $\Delta \delta$ = 6.02\arcsec, but it is just at the edge of the detector, hence it is considered of low priority.

\textbf{HD~88215 (HIP~49809, HR~3991)}: An extended source is present at $\Delta \alpha$ = -7.47\arcsec\ and $\Delta \delta$ = -0.89\arcsec, which is probably a background galaxy.

\textbf{HD~92945 (HIP~52462, V419~Hya)}: The debris disk around HD~92945 has been recently spatially resolved with HST \citep{golimowski2011}. It is not visible in the HiCIAO images.

\textbf{HD~104860 (HIP~58876)}: The only point source in the field of HD~104860 is located at $\Delta \alpha$ = -3.10\arcsec\ and $\Delta \delta$ = 0.55\arcsec, and has already been identified as a background star in \citet{metchev2009}.

\textbf{HD~106591 (HIP~59774, $\delta$~Uma)}: This star was observed in two separate epochs. Two point sources are present in the images. The brighter of the candidates resides at $\Delta \alpha$ = 3.22\arcsec\ and $\Delta \delta$ = -1.25\arcsec\ in the first epoch and $\Delta \alpha$ = 3.08\arcsec\ and $\Delta \delta$ = -1.30\arcsec\ in the second epoch. The fainter one is located at $\Delta \alpha$ = 1.26\arcsec\ and $\Delta \delta$ = -5.57\arcsec\ in the first epoch and $\Delta \alpha$ = 1.06\arcsec\ and $\Delta \delta$ = -5.59\arcsec\ in the second epoch. Neither is therefore physically bound to HD~106591. The brighter candidate however displays a peculiar astrometric behaviour, with a deviation of close to 100~mas from the trajectory of a static background star over a basline of one year. This could imply that it is a field brown dwarf at a similar distance as HD~106591, or otherwise that it is a distant background star with an anomalously high proper motion.

\textbf{HD~107146 (HIP~60074, NLTT~30317)}: The debris disk around HD~107146 has been spatially resolved in the past \citep[e.g.][]{ardila2004,ertel2011}, but since it is smooth and has a nearly face-on orientation, it is not visible in the HiCIAO images. An object is visible at $\Delta \alpha$ = -3.69\arcsec\ and $\Delta \delta$ = -5.07\arcsec, which has been classified as a background galaxy in \citet{ertel2011}.

\textbf{HD~109573 (HR~4796~A, HIP~61498)}: As described in \citet{thalmann2011}, we have spatially resolved the disk in this system using ADI, which enabled us to confirm and strengthen conclusions from previous studies of the system \citep[e.g.][]{schneider1999,schneider2009}, such as the fact that the disk has a non-zero eccentricity. As is also shown in \citet{thalmann2011}, a planet near the gap edge (coplanar with the disk) would have been detectable at a mass of $\sim$3~$M_{\rm Jup}$ at maximum projected separation, but at minimum projected separation the upper limit is much softer ($\sim$17~$M_{\rm Jup}$) due to the relatively high inclination of the target.

\textbf{HD~113337 (HIP~63584, HR~4934)}: There are two epochs of observation available for HD~113337 from HiCIAO, due to the presence of a companion candidate in the data. The candidate has $\Delta \alpha$ = 4.88\arcsec\ and $\Delta \delta$ = -0.10\arcsec\ in the first epoch and $\Delta \alpha$ = 4.97\arcsec\ and $\Delta \delta$ = -0.13\arcsec\ in the second epoch, which demonstrates that it is a background star. Furthermore, archival data from NIRC2 in 2010 places the candidate at $\Delta \alpha$ = 4.74\arcsec\ and $\Delta \delta$ = -0.07\arcsec\, further strengthening this conclusion.

\textbf{HD~128311 (HIP~71395, HN~Boo)}: The single candidate that can be seen in the HiCIAO field at $\Delta \alpha$ = 4.33\arcsec\ and $\Delta \delta$ = -6.38\arcsec\ has been established as a background star in \citet{heinze2010}.

\textbf{HD~139664 (HIP~76829, NLTT~40843)}: A spatially resolved scattered light HST image of the debris disk around HD~139664 exists \citep{kalas2006}. Although the disk has a high inclination, it appears that it was too faint to be detectable in the HiCIAO images.

\begin{figure}[p]
\centering
\includegraphics[width=8cm]{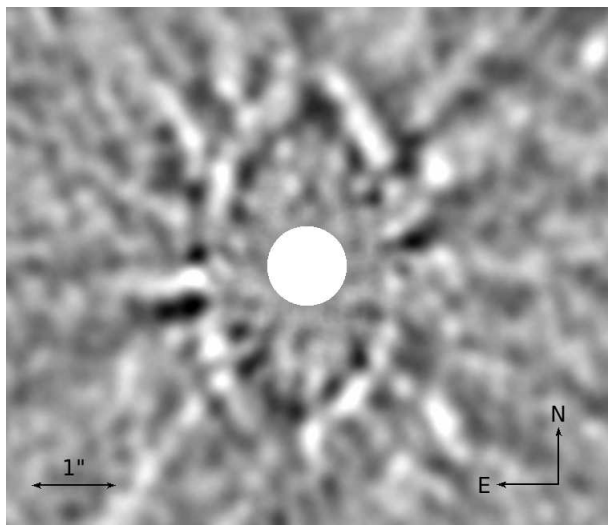}
\caption{Image of the disk around HD~141569. A Gaussian smoothing kernel of 20 pixel FWHM has been applied to the data. Apparent point sources in the image are due to this smoothing. The image is a zoom-in of the central region to more clearly show the disk structure; the binary companion that is present in the full field of view is therefore not visible in this image.}
\label{f:hip77542ring}
\end{figure}

\textbf{HD~141569 (HIP~77542)}: Despite the fact that the disk around HD~141569 is smooth and has a relatively low inclination, it is nonetheless visible in our HiCIAO images (see Fig. \ref{f:hip77542ring}) due to the high surface brightness. As expected, the $S/N$ is lower than in HST images of the target \citep{clampin2003}. For point sources on the other hand, HiCIAO provides strong limits, with sensitivity down to 1~$M_{\rm jup}$ planets in the sensitivity-limited region. The already known binary companion \citep{weinberger2000} is present toward the edge of the field of view.

\textbf{HD~146897 (HIP~79977)}: This USco (Upper Scorpius OB association) member has a debris disk \citep{chen2006}, which was spatially resolved for the first time with HiCIAO, as we reported in \citet{thalmann2013}. The disk has an inner gap within $\sim$40~AU \citep{chen2011}, but owing to the large distance of 123~pc to the target \citep{vanleeuwen2007}, the gap itself cannot be confidently distingished in the existing data, and any giant planet that might be responsible for the gap would have been easily missed, particularly since the disk orientation is close to edge-on.

\textbf{HD~161868 (HIP~87108, $\gamma$~Oph)}:
The star $\gamma$~Oph is in a relatively crowded field with several background stars, although all are outside of 5\arcsec\ separation. The HiCIAO image is relatively shallow and does not reveal as many candidates as archival NICI data from 2009. However, there are two candidates that overlap between the two data sets. One candidate has $\Delta \alpha$ = -6.10\arcsec\ and $\Delta \delta$ = -0.10\arcsec\ in the HiCIAO data and $\Delta \alpha$ = -6.18\arcsec\ and $\Delta \delta$ = -0.29\arcsec\ in the NICI data, and the other has $\Delta \alpha$ = 6.05\arcsec\ and $\Delta \delta$ = 3.89\arcsec\ in the HiCIAO data and $\Delta \alpha$ = 5.95\arcsec\ and $\Delta \delta$ = 3.63\arcsec\ in the NICI data. As expected from the large separations, both candidates are background stars.

\textbf{HD~175742 (HIP~92919, V775~Her)}: There is an object in the field which, judging by its morphology, is probably a close background binary star. It has been observed in two HiCIAO epochs with $\Delta \alpha$ = 1.72\arcsec\ and $\Delta \delta$ = 1.97\arcsec\ in the first epoch and $\Delta \alpha$ = 1.59\arcsec\ and $\Delta \delta$ = 2.24\arcsec\ in the second, confirming its physically unrelated status. In addition, there is an archival epoch from NIRC2 in 2010 where the candidate is located at $\Delta \alpha$ = 1.90\arcsec\ and $\Delta \delta$ = 1.81\arcsec, further strenghtening this conclusion.

\textbf{HD~183324 (HIP~95793, V1431~Aql)}:
The brightest and closest companion candidate to HD~183324 has been observed several times with 8m-class telescopes. In the HiCIAO data, it is located at $\Delta \alpha$ = -0.73\arcsec\ and $\Delta \delta$ = 1.71\arcsec. In archival H-band Keck/NIRC2 images from 2010 it is located at $\Delta \alpha$ = -0.73\arcsec\ and $\Delta \delta$ = 1.63\arcsec. The motion clearly demonstrates that the candidate is a physically unrelated background object. There are also three other candidates inside of 5\arcsec\ in the data; one in the North-East located at $\Delta \alpha$ = 3.25\arcsec\ and $\Delta \delta$ = 1.36\arcsec\ in the HiCIAO image and $\Delta \alpha$ = 3.24\arcsec\ and $\Delta \delta$ = 1.29\arcsec\ in the Keck image, one in the North-West located at $\Delta \alpha$ = 3.40\arcsec\ and $\Delta \delta$ = -1.16\arcsec\ in the HiCIAO image and $\Delta \alpha$ = 3.40\arcsec\ and $\Delta \delta$ = -1.20\arcsec\ in the Keck image, and one towards the South located at $\Delta \alpha$ = 1.49\arcsec\ and $\Delta \delta$ = -4.29\arcsec\ in the HiCIAO image and $\Delta \alpha$ = 1.50\arcsec\ and $\Delta \delta$ = -4.35\arcsec\ in the Keck image. Hence, all of these are also physically unrelated to the target star. 

\textbf{HD~192263 (HIP~99711, V1703~Aql)}: Aside from its debris disk, HD~192263 also hosts a planet candidate detected through radial velocity \citep[e.g.][]{santos2003}. In \citet{chauvin2006}, it is mentioned that several candidates have been discovered and confirmed to be background stars in NACO images of HD~192263. We observe one of these objects within the HiCIAO field of view at $\Delta \alpha$ = -4.41\arcsec\ and $\Delta \delta$ = -5.83\arcsec, and otherwise no new objects.

\begin{figure}[p]
\centering
\includegraphics[width=8cm]{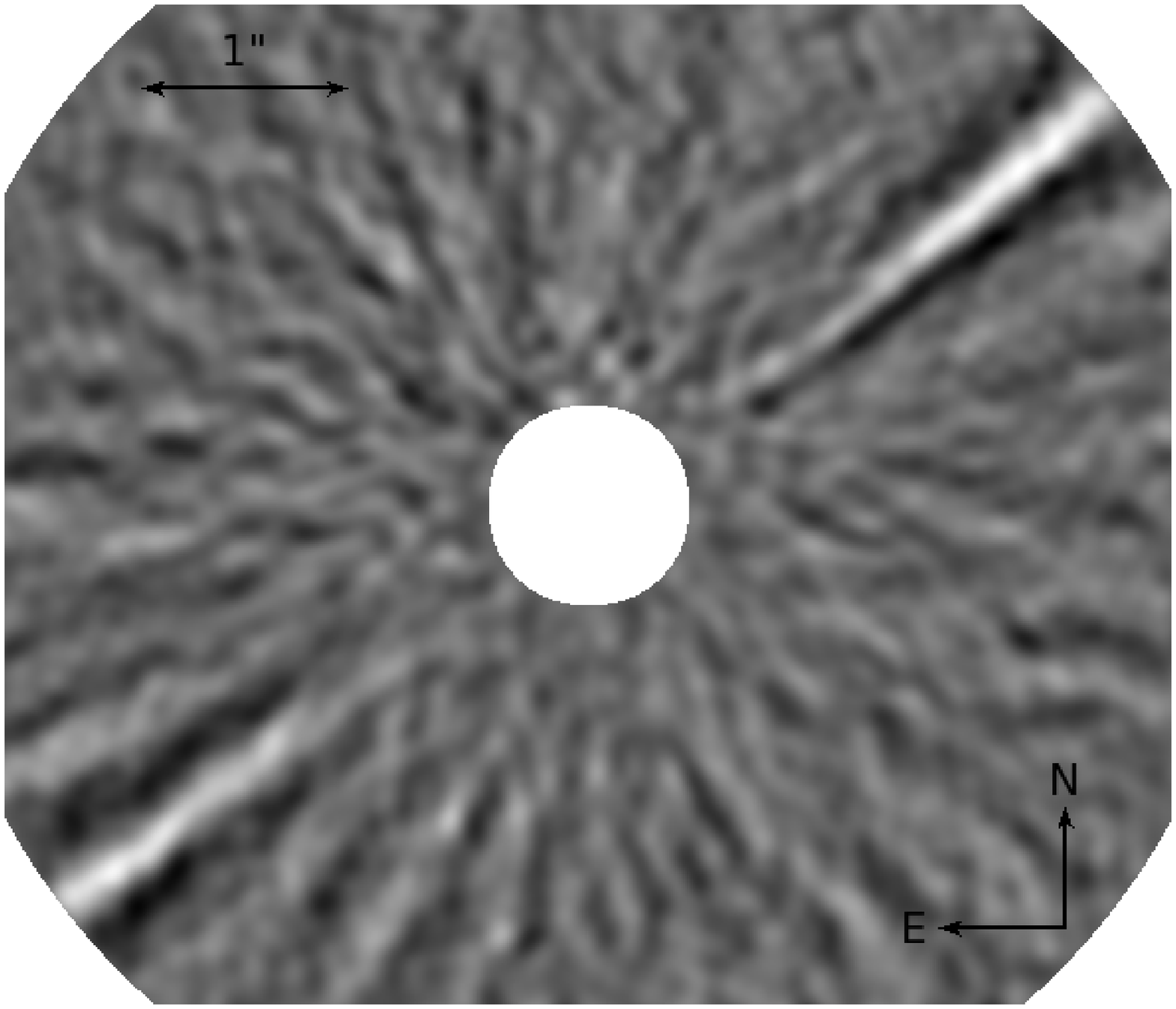}
\caption{Image of the disk around AU~Mic. The characteristic edge-on disk spans diagonally from the South-East to the North-West. The image was acquired with the regular LOCI-based ADI procedure, which causes the black shadows seen on both sides of the disk. A Gaussian smoothing kernel of 10 pixel FWHM has been applied to the data.}
\label{f:aumicedge}
\end{figure}

\textbf{HD~197481 (HIP~102409, AU~Mic)}: Best known as AU~Mic, this star has a well known debris disk which shows up clearly in our data (see Fig. \ref{f:aumicedge}). The field of view is smaller than for most stars in our sample, due to the PDI setting that was used for this observation (see Sect. \ref{s:observations}).

\textbf{HD~206860 (HIP~107350, HN~Peg)}: A candidate is visible at $\Delta \alpha$ = 1.69\arcsec\ and $\Delta \delta$ = 2.45\arcsec. We retrieved the companion in NIRI data from 2006, where the candidate is located at $\Delta \alpha$ = 2.91\arcsec\ and $\Delta \delta$ = 1.93\arcsec, consistent with a background star.

\textbf{HD~207129 (HIP~107649, NLTT~52100)}: Spatially resolved images in scattered light of HD~207129 have been acquired with HST \citep{krist2010}. It is spatially extended and very faint, hence as expected, it is invisible in the HiCIAO images.

\textbf{HD~281691 (V1197~Tau)}: We observe a previously known companion in the HiCIAO images at $\Delta \alpha$ = 4.33\arcsec\ and $\Delta \delta$ = 5.22\arcsec, which was first discovered by \citet{kohler1998} and has been confirmed by \citet{metchev2009}. 


\section{Discussion}
\label{s:discussion}

Both the $\beta$~Pic and HR~8799 systems have debris disks with gaps or cavities in them, and directly imaged planets that are consistent with being responsible for carving these features. Given that we are sensitive to similar mass planets in our observations, and given that we have constraints on the semi-major axis space where the gaps originate from the spectral energy distributions (SEDs) of the targets, it is possible to address to which extent similarly massive planets are responsible for debris disk gaps in general. Given the many caveats involved in such a study however, such an analysis should be treated with caution. 

\begin{table*}[p]
\caption{Target properties}
\label{t:properties}
\centering
\small{
\begin{tabular}{lcccccccccccc}
\hline
\hline
HD ID & SpT & H & Dist & $\tau_{\rm l}$\tablenotemark{a} & $\tau_{\rm u}$\tablenotemark{a} & $\tau$ ref\tablenotemark{b} & $a_{\rm dust}$\tablenotemark{c} & $a$ ref\tablenotemark{b} & $f_{\rm y}$\tablenotemark{d} & $f_{\rm o}$\tablenotemark{d} & $m_{\rm s,y}$\tablenotemark{e} & $m_{\rm s,o}$\tablenotemark{e} \\
 & & (mag) & (pc) & (Myr) & (Myr) & & (AU) & & (\%) & (\%) & ($M_{\rm jup}$) & ($M_{\rm jup}$) \\
\hline
HD 377	&	G2	&	6.15	&	39.1	&	25	&	220	&	A08	&	10	&	H08	&	0.0	&	0.0	&	2	&	6	\\
HD 7590	&	G0	&	5.26	&	23.2	&	420	&	500	&	P09	&	49	&	P09	&	84.1	&	74.8	&	7	&	8	\\
HD 8907	&	F8	&	5.49	&	34.8	&	100	&	400	&	MH09	&	61	&	R07	&	94.4	&	74.6	&	3	&	7	\\
HD 9672	&	A1	&	5.53	&	59.4	&	30	&	50	&	Z12	&	59	&	R07	&	66.1	&	35.0	&	3	&	4	\\
HD 10008	&	G5	&	5.90	&	24.0	&	150	&	300	&	L07	&	9	&	P09	&	0.0	&	0.0	&	3	&	5	\\
HD 12039	&	G4	&	6.56	&	40.9	&	20	&	50	&	Z04	&	8	&	C09	&	0.0	&	0.0	&	1	&	2	\\
HD 15115	&	F2	&	5.86	&	45.2	&	10	&	14	&	M11	&	35	&	R07	&	91.1	&	89.6	&	1	&	1	\\
HD 15745	&	F2	&	6.61	&	63.5	&	10	&	14	&	M11	&	22	&	R07	&	21.7	&	13.0	&	1	&	1	\\
HD 17925	&	K1.5	&	4.23	&	10.4	&	40	&	130	&	L07	&	4	&	H08	&	12.1	&	0.0	&	1	&	2	\\
HD 25457	&	F5	&	4.34	&	18.8	&	50	&	100	&	L06,J07	&	15	&	R07	&	71.2	&	39.2	&	2	&	3	\\
HD 31295	&	A0	&	4.52	&	35.7	&	10	&	100	&	R05,R07	&	47	&	R07	&	64.8	&	0.0	&	3	&	10	\\
HD 40136	&	F2	&	2.99	&	14.9	&	300	&	1410	&	B06,R07	&	6	&	R07	&	0.0	&	0.0	&	4	&	11	\\
HD 60737	&	G0	&	6.31	&	39.3	&	80	&	320	&	C09	&	35	&	C09	&	61.3	&	8.0	&	3	&	6	\\
HD 69830	&	G8	&	4.36	&	12.5	&	5700	&	6100	&	MH08	&	1	&	B11	&	0.0	&	0.0	&	27	&	28	\\
HD 70573	&	G1	&	7.28	&	46.0	&	30	&	125	&	A08	&	28	&	H08	&	71.6	&	4.7	&	2	&	4	\\
HD 72905	&	G1.5	&	4.28	&	14.4	&	50	&	200	&	MH09	&	7	&	H08	&	17.2	&	0.0	&	2	&	4	\\
HD 73350	&	G0	&	5.32	&	24.0	&	370	&	650	&	P09	&	19	&	P09	&	0.0	&	0.0	&	6	&	8	\\
HD 73752	&	G5	&	3.59	&	19.4	&	1600	&	7180	&	M10	&	21	&	R07	&	0.0	&	0.0	&	25	&	53	\\
HD 76151	&	G3	&	4.63	&	17.4	&	1390	&	1890	&	V12	&	6	&	T08	&	0.0	&	0.0	&	14	&	16	\\
HD 88215	&	F2	&	4.46	&	27.7	&	480	&	1760	&	C11	&	5	&	T08	&	0.0	&	0.0	&	9	&	16	\\
HD 91312	&	A7	&	4.06	&	34.6	&	200	&	420	&	R07,V12	&	181	&	R07	&	99.4	&	26.9	&	7	&	10	\\
HD 92945	&	K1.5	&	5.77	&	21.4	&	80	&	120	&	L07	&	24	&	R07	&	60.8	&	36.9	&	4	&	5	\\
HD 102647	&	A3	&	1.92	&	11.0	&	50	&	520	&	R05,R07,V12	&	12	&	R07	&	77.1	&	0.0	&	2	&	7	\\
HD 104860	&	F8	&	6.58	&	45.5	&	20	&	80	&	MH09	&	41	&	H08	&	98.8	&	84.7	&	1	&	2	\\
HD 106591	&	A3	&	3.31	&	24.7	&	300	&	490	&	R05,V12	&	16	&	R07	&	0.0	&	0.0	&	6	&	9	\\
HD 107146	&	G2	&	5.61	&	27.5	&	80	&	200	&	A08	&	27	&	R07	&	74.7	&	46.5	&	2	&	3	\\
HD 109085	&	F2	&	3.37	&	18.3	&	600	&	1300	&	L07	&	5	&	R07	&	0.0	&	0.0	&	10	&	15	\\
HD 109573	&	A0	&	5.79	&	72.8	&	10	&	14	&	Z04	&	33	&	R07	&	32.2	&	19.9	&	1	&	1	\\
HD 110411	&	A0	&	4.76	&	36.3	&	100	&	500	&	R07,V12	&	38	&	R07	&	30.8	&	0.0	&	4	&	10	\\
HD 112429	&	F0	&	4.60	&	29.3	&	50	&	450	&	P09	&	24	&	P09	&	54.7	&	0.0	&	3	&	9	\\
HD 113337	&	F6	&	5.05	&	36.9	&	20	&	60	&	M11	&	18	&	R07	&	41.3	&	0.0	&	1	&	3	\\
HD 125162	&	A0	&	4.03	&	30.4	&	180	&	320	&	R05,R07,V12	&	33	&	R07	&	0.0	&	0.0	&	7	&	9	\\
HD 127821	&	F4	&	5.10	&	31.8	&	170	&	270	&	M11	&	56	&	R07	&	71.0	&	47.4	&	6	&	7	\\
HD 128167	&	F2	&	3.46	&	15.8	&	1000	&	4780	&	R07,V12	&	90	&	R07	&	45.7	&	0.0	&	10	&	25	\\
HD 128311	&	K0	&	5.30	&	16.5	&	140	&	460	&	M10	&	5	&	T08	&	0.0	&	0.0	&	2	&	5	\\
HD 135599	&	K0	&	5.12	&	15.8	&	190	&	230	&	P09	&	11	&	L09	&	68.7	&	64.7	&	2	&	3	\\
HD 139006	&	A0	&	2.39	&	23.0	&	270	&	500	&	R05,R07,V12	&	17	&	R07	&	0.0	&	0.0	&	4	&	7	\\
HD 139664	&	F4	&	3.73	&	17.4	&	150	&	300	&	L06	&	25	&	R07	&	57.3	&	27.7	&	4	&	6	\\
HD 141569	&	B9.5	&	6.86	&	116.1	&	4	&	5	&	M04	&	29	&	R07	&	46.7	&	2.1	&	1	&	1	\\
HD 146897	&	F2	&	7.85	&	122.7	&	5	&	10	&	Z99,P12,S12	&	16	&	C06	&	0.0	&	0.0	&	1	&	2	\\
HD 152598	&	F0	&	4.54	&	29.2	&	140	&	280	&	M09	&	9	&	R07	&	0.0	&	0.0	&	5	&	7	\\
HD 161868	&	A0	&	3.66	&	31.5	&	180	&	310	&	R05,R07	&	59	&	R07	&	65.7	&	14.4	&	7	&	9	\\
HD 162917	&	F4	&	4.83	&	31.1	&	200	&	800	&	R07	&	21	&	R07	&	0.0	&	0.0	&	6	&	12	\\
HD 175742	&	K0	&	5.76	&	21.4	&	40	&	60	&	P09	&	4	&	P09	&	0.0	&	0.0	&	1	&	2	\\
HD 183324	&	A0	&	5.59	&	61.2	&	5	&	20	&	R05	&	18	&	MW09	&	3.7	&	0.0	&	1	&	2	\\
HD 192263	&	K2.5	&	5.69	&	19.3	&	550	&	570	&	S05	&	11	&	D11	&	0.0	&	0.0	&	6	&	6	\\
HD 197481	&	M1	&	4.83	&	9.9	&	10	&	14	&	P09	&	10	&	R07	&	94.7	&	94.7	&	1	&	1	\\
HD 206860	&	G0	&	4.60	&	17.9	&	150	&	300	&	L07	&	7	&	T08	&	0.0	&	0.0	&	3	&	4	\\
HD 207129	&	G0	&	4.31	&	16.0	&	600	&	3200	&	R07,MH08	&	28	&	R07	&	34.9	&	0.0	&	8	&	19	\\
HD 281691	&	G8	&	8.51	&	73	&	10	&	30	&	M08	&	23	&	C09	&	74.4	&	24.8	&	2	&	2	\\
\hline
\end{tabular}
}
\tablenotetext{a}{\scriptsize{The lower and upper limits on the age are denoted $\tau_{\rm l}$ and $\tau_{\rm u}$.}}
\tablenotetext{b}{\scriptsize{The references are abbreviated as follows: A08: \citep{apai2008}, B06: \citep{beichman2006}, C06: \citep{chen2006}, C09: \citep{carpenter2009}, C11: \citep{casagrande2011}, D11: \citep{dodson2011}, H08: \citep{hillenbrand2008}, J07: \citep{janson2007}, L06: \citep{lopez2006}, L07: \citep{lafreniere2007b}, L09: \citep{lawler2009}, M04: \citep{merin2004}, M08: \citep{meyer2008}, MH08: \citep{mamajek2008}, M09: \citep{moor2009}, MH09: \citep{metchev2009}, MW09: \citep{morales2009}, M10: \citep{maldonado2010}, M11: \citep{moor2011}, P09: \citep{plavchan2009}, P12: \citep{pecaut2012}, R05: \citep{rieke2005}, R07: \citep{rhee2007}, S05: \citep{saffe2005}, S12: \citep{song2012}, T08: \citep{trilling2008}, V12: \citep{vican2012}, Z99: \citep{dezeeuw1999}, Z04: \citep{zuckerman2004}, Z12: \citep{zuckerman2012}.}}
\tablenotetext{c}{\scriptsize{Location of the dust, see text for discussion.}}
\tablenotetext{d}{\scriptsize{Detection probability for a 10$M_{\rm jup}$ planet at semi-major axis $a_{\rm dust}$, denoted $f_{\rm y}$ for the youngest and $f_{\rm o}$ for the oldest age.}}
\tablenotetext{e}{\scriptsize{Mass limits in the sensitivity-limited regime are denoted $m_{\rm s,y}$ for the youngest and $m_{\rm s,o}$ for the oldest age.}}
\end{table*}

One primary issue in the analysis is the uncertainty in the location of the gap. It is possible to constrain the spatial distribution of the circumstellar dust from the SED by constraining the temperature, but since only a very limited number of data points are available in general, there are ambiguities between the location and the radiative properties of the dust. In this study, we adopt values of $a_{\rm dust}$ from the literature based on the global assumption that the dust grains emit like blackbodies. How the resulting physical separation relates to the semi-major axis of a given hypothetical shepherding planet in the system is another complex uncertainty. Here, we simply take the $a_{\rm dust}$ itself to represent the physical scale around which we wish to evaluate the presence or absence of a planet; the motivation being that the dominating disk flux should arise close to the inner edge (since that is where the dust is hottest and, in general, most dense) and that the planet responsible for carving the gap should be close to the edge. This is not necessarily relevant if, for instance, there are multiple planets responsible for the gap. With regards to planet detectability near the gap, the gap locations adopted here are probably very conservative, as can be seen in \citet{booth2013}. In all cases studied by \citet{booth2013} where the real gap location could be observed, the real location is never smaller than the blackbody prediction, but is often larger by a factor 2.

\begin{figure}[p]
\centering
\includegraphics[width=8cm]{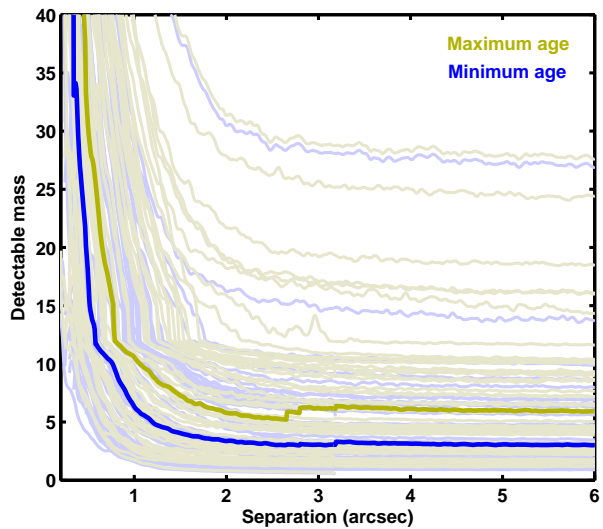}
\caption{Detectable mass as a function of angular separation, based on the COND/DUSTY models. Blue curves correspond to the mass at the lower limit of the age range estimated for each target, and gold curves correspond to the mass at the upper limit. The thick opaque lines are the median mass detection limits across the sample, and the lighter narrower lines are the individual cases.}
\label{f:dd_mcurves}
\end{figure}

We derive mass detection limits from the contrast curves using COND- and DUSTY-based evolutionary models \citep{chabrier2000,allard2001,baraffe2003} and the age limits in Table \ref{t:properties} (see Fig. \ref{f:dd_mcurves}). COND was used whenever the predicted tempearture was below 1700~K, and DUSTY when it was above this limit. These `hot-start' models may over-predict the brightness for a given mass and age if the initial entropy is lower than assumed in those models \citep[see e.g.][]{spiegel2012}. However, the exoplanets that have been discovered to date are consistent with hot-start conditions and exclude at least the coldest ranges of initial conditions \citep[e.g.][]{janson2011,bonnefoy2013,marleau2013}. Furthermore, the absence of heating from deuterium burning in the COND/DUSTY models may conversely under-predict the brightness for a given mass and age \citep{molliere2012}. Nonetheless, the uncertainties in mass-luminosity relationships is a further uncertainty that should be kept in mind.

\begin{figure}[p]
\centering
\includegraphics[width=8cm]{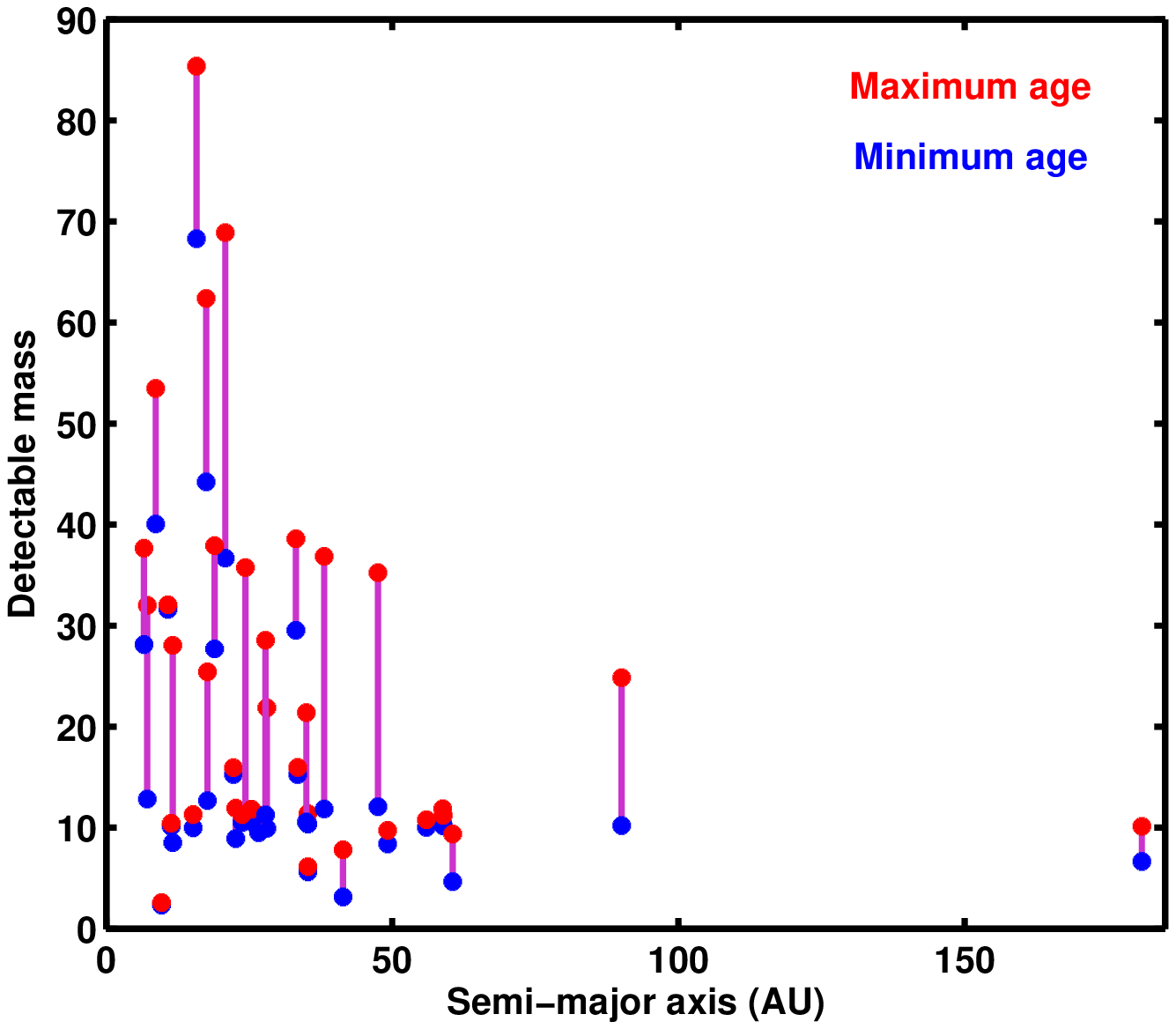}
\caption{Detectable mass and semi-major axis at the estimated gap edge of the debris disks. Blue points are the masses at the lower age limit, and red points are the masses at the upper limit.}
\label{f:dd_edgemass}
\end{figure}

In order to put the issue of gap-opening super-Jupiters in a statistical context, we evaluate the probability that planets with masses similar to that of $\beta$~Pic~b of $\sim$10~$M_{\rm jup}$ \citep[][and Currie et al. in prep.]{bonnefoy2013} would be detectable near the gap in each observed system. This is done by calculating the full projected separation distribution corresponding to a semi-major axis of $a_{\rm dust}$ for random orbital orientations and a uniform eccentricity distribution between 0.0 and 0.6 \citep{janson2011,bonavita2012}. The fraction of 10~$M_{\rm jup}$ planets that are detectable in a given system is denoted $f_{\rm y}$ for the lower limit of the age of the star and $f_{\rm o}$ for the upper limit. The individual values of $f_{\rm y}$ and $f_{\rm o}$ are listed in Table \ref{t:properties}. In some of the systems such a planet is simply not detectable (0\% in both $f_{\rm y}$ and $f_{\rm o}$), while in the best cases the fraction is close to 100\%. From the collection of these values and the fact that no planets were detected in the sample, we can estimate an upper limit on the frequency of planets with equal or higher mass than $\beta$~Pic~b near the estimated gap edge, using Bayes theorem following the procedure in \citet{janson2011}. As a result, we find that at 95\% confidence, $<$15.2\% of the stars host such planets in the extreme case where the younger age limit is adopted in all cases, and $<$30.1\% in the opposite case where the upper limits are adopted. In other words, if giant planets are a dominant cause of gaps in debris disks, the majority of them must be less massive than $\beta$~Pic~b. In either case, it implies that $\beta$~Pic~b is probably in the upper mass range of any gap-causing planets that may exist. 

An illustration of typical mass detection limits around $a_{\rm dust}$ for the individual stars is shown in Fig. \ref{f:dd_edgemass}, where the detectable mass is evaluated at $\alpha _{\rm dust} = a_{\rm dust}/d/1.26$, which represents the average angular separation of a planet with semi-major axis $a_{\rm dust}$ for random orbital orientations \citep{fischer1992}. Here, $d$ denotes the distance to the target. The mass limits in our survey are contrast-limited rather than sensitivity-limited, hence it would be possible to substantially enhance the limits with upcoming Extreme Adaptive Optics-assisted instruments such as SPHERE, GPI, or CHARIS \citep{beuzit2008,macintosh2008,peters2012}. These facilities may thus be able to detect a large number of gap-opening super-Jupiters if they are relatively common, or otherwise put yet more stringent limits on their presence and properties.

\section{Conclusions}
\label{s:conclusions}

In this study, we have presented high-contrast imaging of a sample of 50 stars primarily in the G--A-type range with known infrared excess due to debris disks, using the HiCIAO camera at the Subaru telescope. Targets were particularly selected for if they had excess only at long wavelengths, implying cold debris disks with an inner gap, possibly carved out by massive planets within the disk. The targets were observed both in order to attempt to spatially resolve the disk, as well as to try to detect the putative planets that may be responsible for the disk morphology. No planets were discovered, despite the fact that $\beta$~Pic~b-like planets ($\sim$10~$M_{\rm jup}$) could have been detected near the esimated gap edges in many cases. This led to an upper limit of 15--30\% on the frequency of such planets, implying that if planets are a general cause of the commonly existing gaps in debris disk systems, then they must generally be lower in mass than $\beta$~Pic~b. Five deris disks have been spatially resolved during the survey, two of which have already been presented in previous publications \citep{thalmann2011,thalmann2013}. Future studies with upcoming instrumentation will be able to put yet more stringent constraints on planet occurrences in debris disk systems, by probing down to smaller planetary masses and smaller semi-major axes, and thus may conclusively address whether the gaps in debris disks are typically caused by planets, or whether other mechanisms dominate the disk architecture.

\acknowledgements
Support for this work was provided by NASA through Hubble Fellowship grant HF-51290.01 awarded by the Space Telescope Science Institute, which is operated by the Association of Universities for Research in Astronomy, Inc., for NASA, under contract NAS 5-26555. J.C. was supported by NSF award 1009203.  Archival data from the Subaru, Gemini, Keck and Hubble telescopes have been used as part of this study. We acknowledge the cultural significance of Mauna Kea to the indigenous population of Hawaii. This study made use of the CDS services SIMBAD and VizieR, as well as the SAO/NASA ADS service.

\clearpage

\end{document}